\documentclass[aps,prl,twocolumn,groupedaddress,showpacs]{revtex4-1}

\usepackage{amssymb}
\usepackage{color,graphicx}

\begin{document}

\title{Observation of Solitonic Vortices in Bose-Einstein Condensates}

\author{Simone Donadello$^{1}$}
\author{Simone Serafini$^{1}$}
\author{Marek Tylutki$^{1}$}
\author{Lev P. Pitaevskii$^{1,2}$}
\author{Franco Dalfovo$^{1}$}
\author{Giacomo Lamporesi$^{1}$}
\author{Gabriele Ferrari$^{1}$}
\email[]{ferrari@science.unitn.it}

\affiliation{\textit{1 INO-CNR BEC Center and Dipartimento di Fisica, Universit\`a di Trento, 38123 Povo, Italy\\
2 Kapitza Institute for Physical Problems RAS, Kosygina 2, 119334 Moscow, Russia}}
\date{\today}

\begin{abstract}
We observe solitonic vortices in an atomic Bose-Einstein condensate after free expansion. Clear signatures of the nature of such defects are the twisted planar density depletion around the vortex line, observed in absorption images, and the double dislocation in the interference pattern obtained through homodyne techniques. Both methods allow us to determine the sign of the quantized circulation. Experimental observations agree with numerical simulations. These solitonic vortices are the decay product of phase defects of the BEC order parameter spontaneously created after a rapid quench across the BEC transition in a cigar-shaped harmonic trap and are shown to have a very long lifetime.
\end{abstract}


\maketitle

In a recent experiment \cite{LamporesiKZM13}, we reported on the observation of defects in a Bose-Einstein condensate (BEC) after rapid quench across the BEC transition. Those defects showed a planar structure, as expected for grey solitons spontaneously created {\it via} the Kibble-Zurek mechanism (KZM) \cite{Kibble76,Zurek85} in an elongated cigar-shaped condensate \cite{Zurek09}, but with a surprisingly long lifetime compared to the one predicted for this type of excitations \cite{Feder00}.
Long living solitonic structures were also observed in superfluid Fermi gases \cite{Yefsah13}. Motivated by these facts, we decided to set up a new protocol for the observation of defects in our BEC in order to better characterize their density and phase structure. In this way, we unambiguously identify them as solitonic vortices, which are likely the remnants of the decay processes of grey solitons. These topological defects were predicted more than a decade ago \cite{Brand02,Komineas03}. Indirect evidences were recently reported in \cite{Becker13}, by looking at collisions between defects produced by phase imprinting techniques. Observations of solitonic vortices were reported also in superfluid Fermi gases in \cite{Ku14} by using a tomographic imaging technique and by tracking the in-trap oscillations of the associated density depletion.
Our results are timely and interesting in view of the ongoing discussions on the nature and the dynamical evolution of vortices, vortex rings and solitons in bosonic \cite{Becker13,Reichl13} and fermionic \cite{Ku14,Bulgac14,Wlazlowski14,Scherpelz14} gases and for their implications in studies of superfluidity and quantum turbulence in a wider context which includes superfluid helium, neutron stars, polariton condensates, and cosmological models.

The experimental apparatus is described in \cite{Lamporesi13}. As in \cite{LamporesiKZM13}, we produce ultracold sodium samples confined in a harmonic axisymmetric magnetic trap with frequencies $\omega_{x}/2\pi=13\,\text{Hz}$ and $\omega_{\perp}/2\pi=131\,\text{Hz}$ ($\omega_y=\omega_z=\omega_{\perp}$). The weak axis of the trap lies along the $x$ direction, in the horizontal plane. We perform RF-forced evaporative cooling using linear ramps across the BEC transition with variable rate. Sufficiently fast ramps induce a temperature quench in the atomic gas triggering the KZM. We choose a quenching ramp of $320$~kHz/s for the data reported in this Letter, in order to produce about one or two defects on average. At the transition ($T\simeq700$ nK) the system contains about $2.5\times10^7$ atoms, leading to cigar-shaped condensates of $1.0\times10^7$ atoms with negligible thermal component after further evaporation and a chemical potential $\mu_{\mathrm{exp}}= k_B (170\,\mathrm{nK}) = 27 \,\hbar\omega_{\perp}$. Atoms are released from the trap and let expand in presence of a vertical magnetic field gradient to compensate for gravity. A long time of flight can then be achieved to clearly observe the defect structure in the condensate, without optical density saturation.

\begin{figure*}[!hbt]
\centering
\includegraphics[width=2.05\columnwidth]{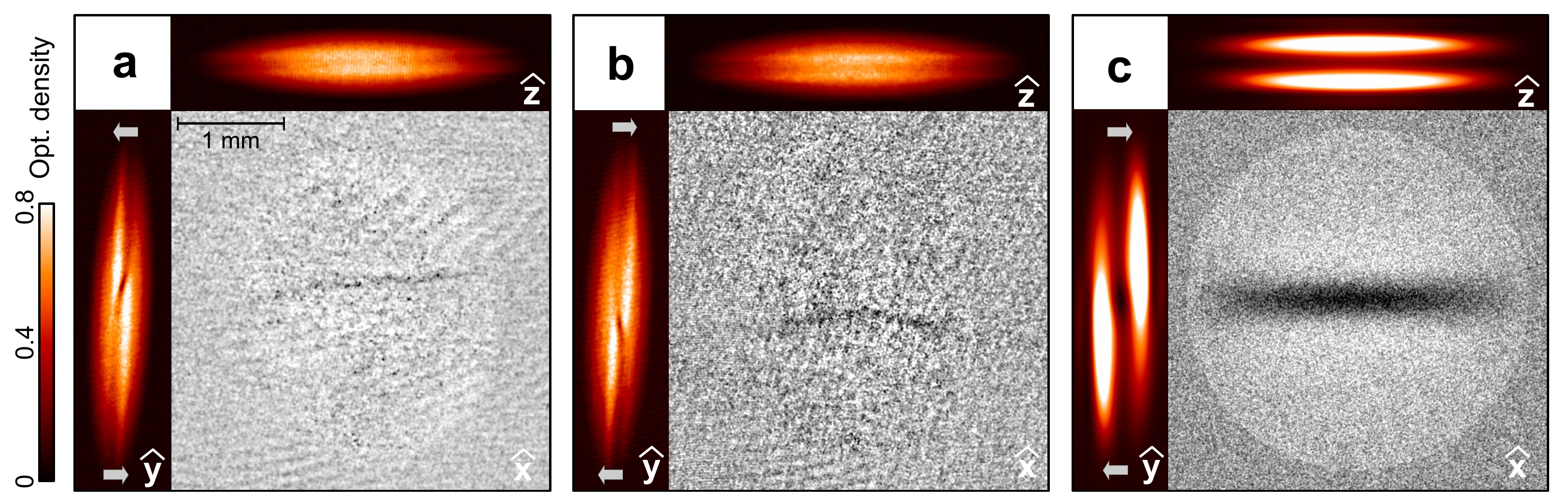}
\caption{(color online) Integrated triaxial density distribution of a BEC after a time of flight of 120 ms in presence of a solitonic vortex aligned along $y$. For each condensate we report absorption images along the two radial directions, horizontal $y$ (left) and vertical $z$ (top), and the residuals of axial ($x$) imaging after subtracting the Thomas-Fermi profile fit. a-b) Experimental snapshots of two condensates with opposite circulation. c) Theoretical 3D calculation with clockwise circulation and a $\mu_{\mathrm{theo}}\simeq\mu_{\mathrm{exp}}/3$ (white noise has been added in order to better compare the theoretical calculation to the experimental results). Arrows indicate the atomic flow. Other examples of solitonic vortices with different orientation and shape can be seen in the Supplemental Material \cite{SupplementalMaterial1}.}
\label{fig:Figure1}
\end{figure*}

\begin{figure}[b!]
\centering
\includegraphics[width=1\columnwidth]{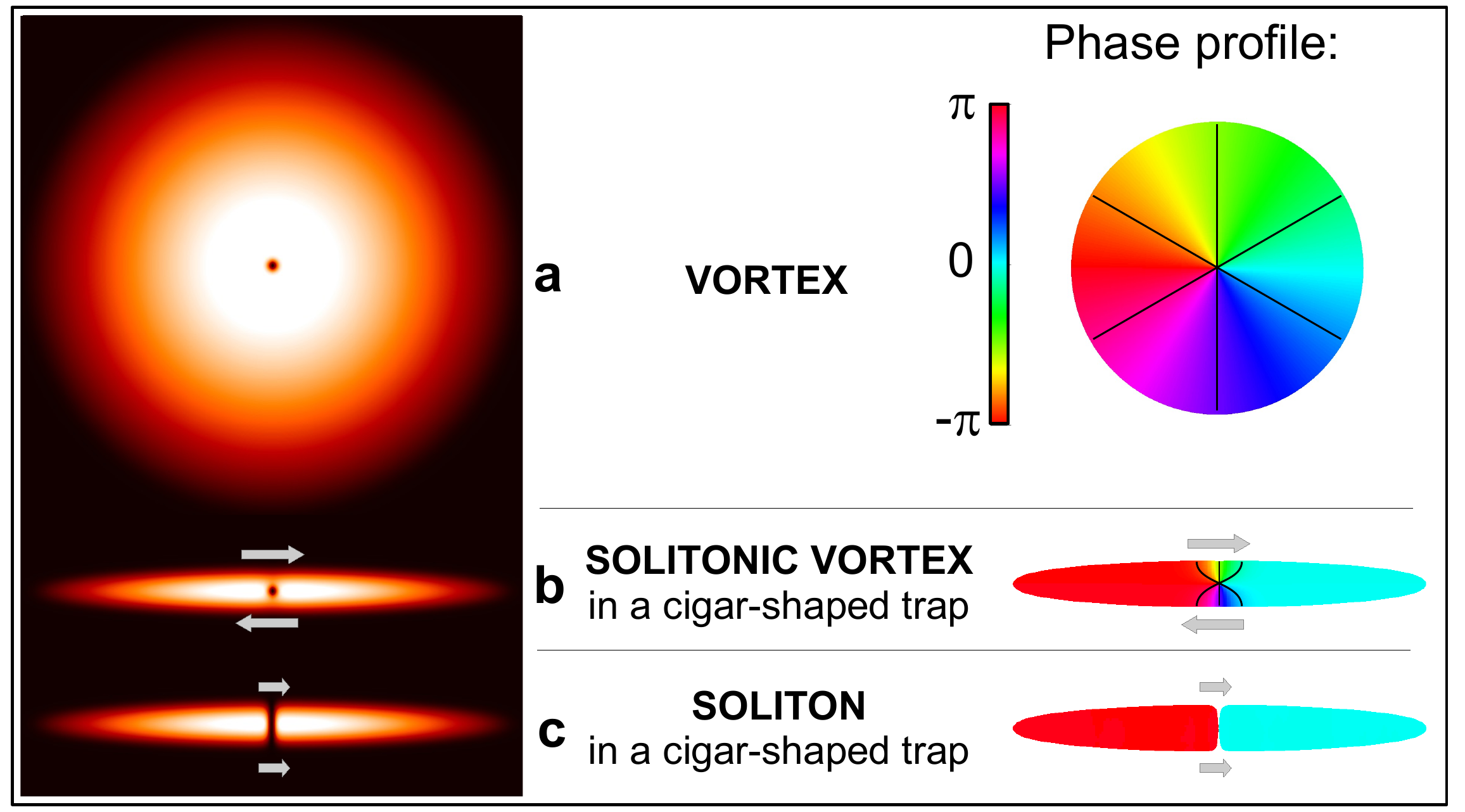}
\caption{(color online) Theoretical solutions of \emph{in-situ} density and phase profiles for: (a) a vortex aligned along the axis of an axially symmetric potential, a solitonic vortex (b) and a soliton (c) oriented perpendicularly to the axis of an elongated axially symmetric trap. Arrows indicate the atomic flow.}
\label{fig:Figure2}
\end{figure}

Fig.~\ref{fig:Figure1}a-b show two typical examples of simultaneous triaxial absorption imaging after a time of flight $t_{\mathrm{TOF}}=120\;\text{ms}$. The expanded condensate has an oblate ellipsoidal shape. For each of the two condensates we report standard absorption imaging along two radial directions $y$ and $z$, and also the residuals of the axial absorption imaging, after the removal of fitting Thomas-Fermi (TF) density profile. Clear vortex lines are visible from the axial direction. The contrast of the vortex lines is low, thus digital filtering of the images was performed to enhance the visibility of the defects.  The anisotropy in the radial confinement is negligible ($< 10^{-4}$) preventing any preferential vortex line orientation in the $yz$ plane. In Fig.~\ref{fig:Figure1} we report two selected images of condensates with a vortical line aligned along $y$ and compare them to a theoretical simulation (Fig.~\ref{fig:Figure1}c); more images are given in the Supplemental Material \cite{SupplementalMaterial1}. In general, any time we see a vortex line from the axial direction we also detect a stripe when looking from both radial directions. This is the first signature of the solitonic nature of the defect. A vortex aligned along the symmetry axis of a cylindrically symmetric trap would exhibit a phase $\phi$ of the order parameter linearly growing from $0$ to $2\pi$ around the axis and both the density profile and the velocity field would be isotropic (Fig.~\ref{fig:Figure2}a). Conversely, if the vortex is oriented along the radial direction of an elongated trap, as in our case, the equi-phase surfaces, which originate from the vortex and must be orthogonal to the condensate outer surface, are forced to bend and to gather in the short radial direction (Fig.~\ref{fig:Figure2}b). The phase gradient is then concentrated in a planar region whose width is much smaller than the axial length of the condensate.
The high phase gradient corresponds to a high atomic velocity field and the system tends to reduce its energy by depleting the density in the plane. The resulting structure is a solitonic vortex \cite{Brand02,Komineas03}. Far from the density depleted region the phase pattern is similar to that of a grey soliton (Fig.~\ref{fig:Figure2}c), the two sides of the condensate having different phases ($\Delta\phi=\pi$ in case of a stationary solitonic vortex and a dark soliton). A key difference is that in a soliton the phase gradient is the same in the whole nodal plane, while in a solitonic vortex it has opposite sign in the two half-planes separated by the vortex line. A consequence is that, if a condensate containing a soliton is released from the trap, the overall density structure is preserved, while in the case of a solitonic vortex the asymmetric flow induces the two density depleted sides to twist in opposite directions. From the standard definition of superfluid velocity, $v=(\hbar/m)\nabla{\phi}$, it follows that the atomic flux is oriented towards increasing phases, hence the depleted regions move in the opposite direction. If at least one extreme of the vortex line is oriented along $y$ or $z$ a remarkable structure appears (see Fig.~\ref{fig:Figure1}): a solitonic plane twisted around a hollow vortex core. Such a feature was already clearly visible in \cite{LamporesiKZM13}. The sign of a vortex, which was previously observed exciting collective modes \cite{Chevy00}, following the vortex core precession \cite{Anderson2000,Freilich10} or using interferometric techniques \cite{Chevy01}, is here extracted from the twist orientation in single absorption images.

\begin{figure*}[t!]
\centering
\includegraphics[width=2.05\columnwidth]{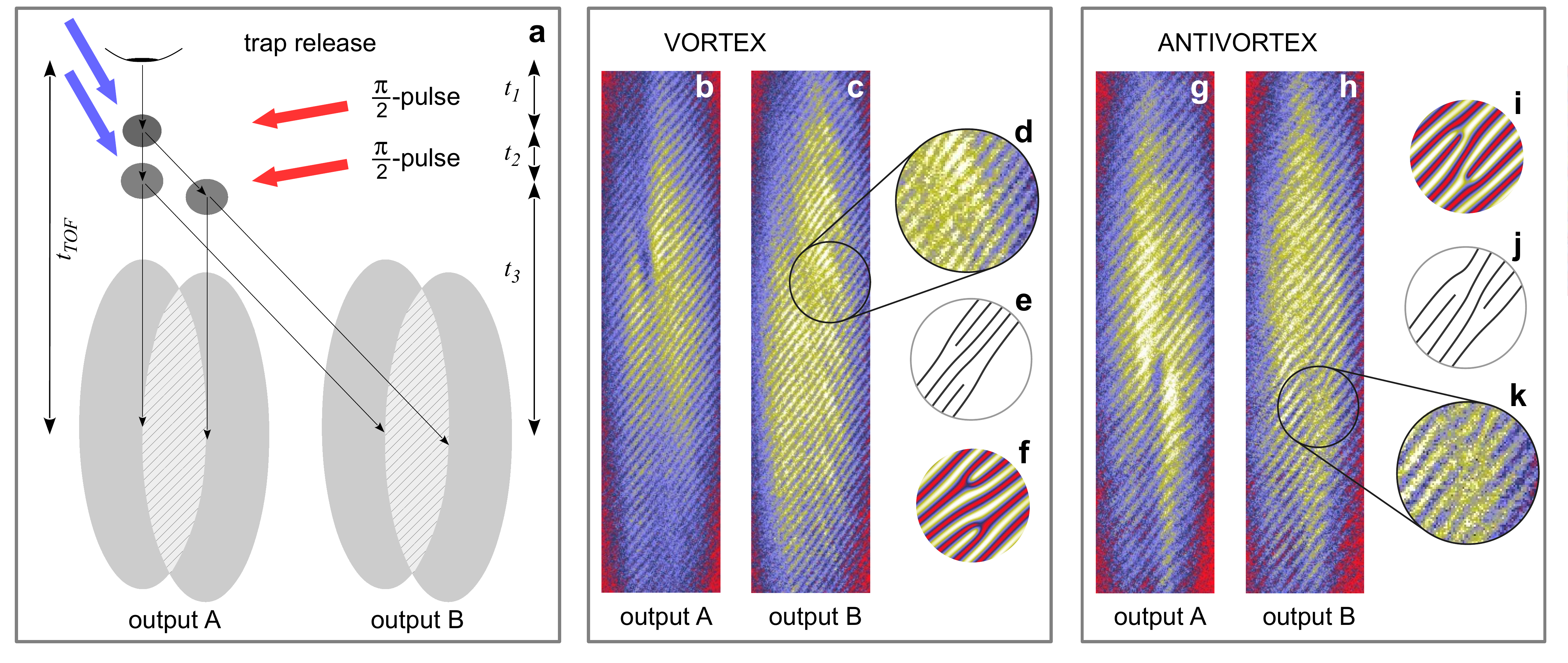}
\caption{(color online) a) Time sequence for the two-pulse Bragg interferometer. After $t_1$ from the atoms release a first $\frac{\pi}{2}$ Bragg pulse splits the condensate in two parts, one at rest and the other moving at the Bragg velocity. After $t_2$ a second pulse further splits each condensate and imaging after a long $t_3$ allows for the separation of the two outputs and for the creation of the interference pattern. b-c) Experimental images of the two interferometer outputs in presence of a vortex with clockwise circulation. d) Zoom on the interference pattern showing the dislocation-antidislocation pair and a guide to the eye (e) sketching the interference pattern. f) 2D numerical simulation of the experiment. g-k) The same, but for an antivortex with counterclockwise circulation.}
\label{fig:Figure3}
\end{figure*}

We compare our observations with the solution of the Gross-Pitaevskii (GP) equation $i \hbar \partial_t \psi = -\frac{\hbar^2}{2m} \nabla^2 \psi + \frac12 m \omega_x^2 (x^2 + \alpha^2 r^2) \psi + g |\psi|^2 \psi$, where $\alpha = \omega_{\perp} / \omega_x$ and $g = 4 \pi \hbar^2 a_s / m$ with $a_s$ being the scattering length. The trapped stationary state is obtained by means of an imaginary time evolution of the GP equation, where the phase pattern of a solitonic vortex is initially imprinted~\cite{Brand02,Komineas03,ParkerPhDthesis}. This state is then used as initial condition for the simulation of the free expansion. In order to speed up the simulation time and make the calculation feasible also for long expansions, we dynamically rescale the GP equation by means of the TF scaling {\it ansatz} in the transverse direction
~\cite{Castin96,Kagan96,Dalfovo97,Modugno03,Massignan03,Tozzo04,Dalfovo99}. A typical result is shown in Fig.~\ref{fig:Figure1}c for a three-dimensional (3D) simulation of a condensate expanding for $120$~ms with $\omega_{\perp} = 10\, \omega_x$ and $\mu_{\mathrm{theo}} = 10\, \hbar \omega_{\perp}$. The expanding solitonic vortex clearly develops a planar density depletion, which twists near the vortex core due to the phase gradient, in qualitative agreement with the experimental data in Fig.~\ref{fig:Figure1}a-b. The width of the depletion region in the simulation is larger, since the chemical potential is about $1/3$ of the experimental one and the corresponding value of the healing length of the condensate is larger \cite{note1}. We notice that the solitonic plane twists during the expansion, but the twist displacement saturates when the mean-field interaction lowers and the expansion proceeds in a ballistic way. We tested this mechanism also by comparing simulations in 2D and 3D and seeing that in 2D the twist is more pronounced and lasts longer \cite{SupplementalMaterial1}. This is consistent with the fact that in 3D the density decreases faster than in 2D as it expands also in the third dimension, and hence the system enters the ballistic regime earlier.

In order to prove the quantized vorticity of the observed defects, we implement a matter wave interferometer \cite{Bolda98,Inouye01,Chevy01} (see Fig.~\ref{fig:Figure3}a). The presence of vorticity appears as a dislocation in the fringe pattern in a heterodyne interferometer \cite{Inouye01}, while in the case of homodyne detection a single vortex gives rise to a pair of dislocations with opposite orientations \cite{Chevy01}. Our interferometer is based on homodyne detection: the original condensate is coherently split in two clouds using coherent Bragg pulses as in \cite{Kozuma99}. Two off-resonant laser beams, whose relative detuning is set by acousto-optics modulators, impinge on atoms with a relative angle of about $110\,^{\circ}$. They propagate in the $xz$ plane and are both linearly polarized along $y$. An optimal detuning of $67\;\text{kHz}$ is needed to excite the first-order Bragg process. Half population transfer ($\frac{\pi}{2}$ pulse) is obtained with an intensity of $12$~mW/cm$^2$ on each beams and a pulse duration of $8\;\mu\text{s}$. The interferometer is an open-type $\frac{\pi}{2}-\frac{\pi}{2}$ one. A first Bragg pulse coherently splits the initial wave-function in two, creating a traveling copy along the direction of the exchanged momentum. The first pulse is applied at $t_1=20\;\text{ms}$ after the release from the trap. In-trap atomic density is too high and interaction effects between the copies would dominate in case of smaller $t_1$. The second Bragg pulse is applied $t_2=1.5\;\text{ms}$ after the first one, when the two copies of the condensate are separated by $d\approx72\;\mu\text{m}$. After the second pulse four equally populated copies of the original wave-function are present, two of them at rest, whereas the others traveling at the Bragg velocity ($4.8\;\text{cm}/\text{s}$). We image the resulting atomic distribution after $t_3=98.5\;\text{ms}$, corresponding to a total expansion time $t_{\mathrm{TOF}}=120\;\text{ms}$. Copies with different velocities become well separated in space, while those with the same velocity spatially overlap and show interference patterns in the density profile. The fringe spacing is given by $\lambda=\frac{h\,t_3}{m\,d}\approx25\;\mu\text{m}$. Fig.~\ref{fig:Figure3}b-c show the two outputs of the Bragg interferometer obtained in presence of a solitonic vortex with clockwise circulation. The hollow vortex core and the associated twist is best visible in output A. In correspondence to the vortex core a dislocation in the interferometric fringes appears (see zoom in Fig.~\ref{fig:Figure3}d-e) together with its opposite counterpart, shifted towards bottom right because of the momentum transfer given by the geometrical Bragg beam setting. In Fig.~\ref{fig:Figure3}f a 2D numerical calculation is reported for comparison. Figures~\ref{fig:Figure3}g-k report the opposite case of a solitonic antivortex, with counterclockwise circulation. These observations confirm that the defects observed in Fig.~\ref{fig:Figure1} are solitonic vortices and also demonstrate the possibility to directly measure the circulation sign from the density profile after expansion.

In the light of these results we can better understand the long lifetime of the defects we observed in our previous work \cite{LamporesiKZM13}.
Assuming that planar solitons are most likely created in the elongated trap {\it via} the KZM during the rapid quench through the BEC transition, they may quickly decay due to a snaking instability. On the basis of GP simulations \cite{Becker13,Shomroni09} and analytic results in the hydrodynamic regime \cite{Kamchatnov08,Cetoli13}, one can roughly estimate the timescale of this decay process in our geometry to range from a few ms to a few tens of ms, for solitons with randomly distributed initial phase and amplitude fluctuations.  As described in the GP simulations of Ref.~\cite{Becker13}, the first stage of the decay can be a vortex ring, subsequently evolving into a solitonic vortex, which has lower energy \cite{Brand02,Komineas03} and whose lifetime is much longer, being topologically protected.

\begin{figure}[t!]
\centering
\includegraphics[width=1.05\columnwidth]{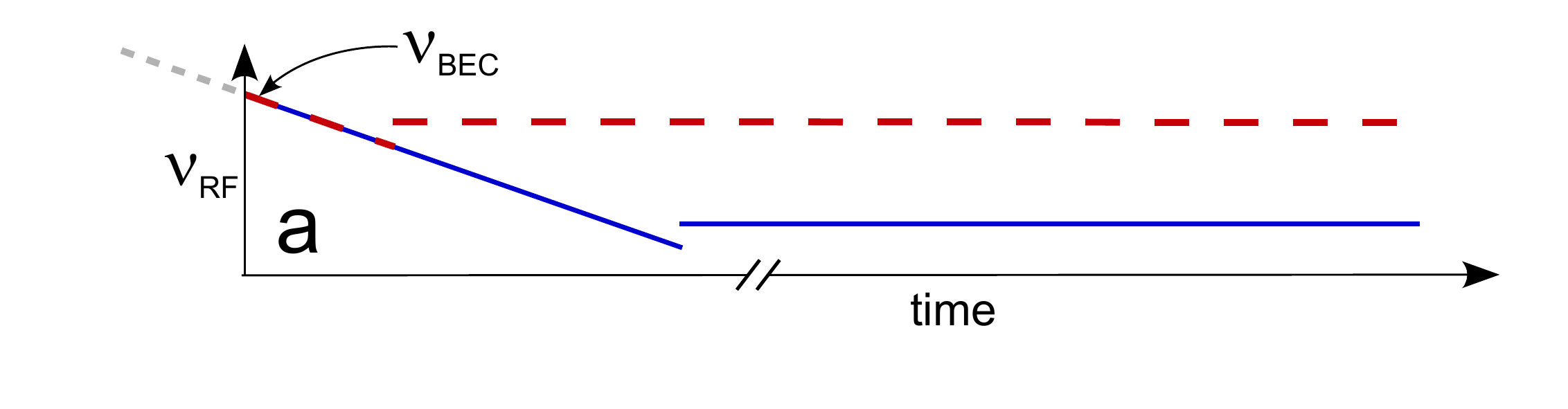}
\includegraphics[width=1.05\columnwidth]{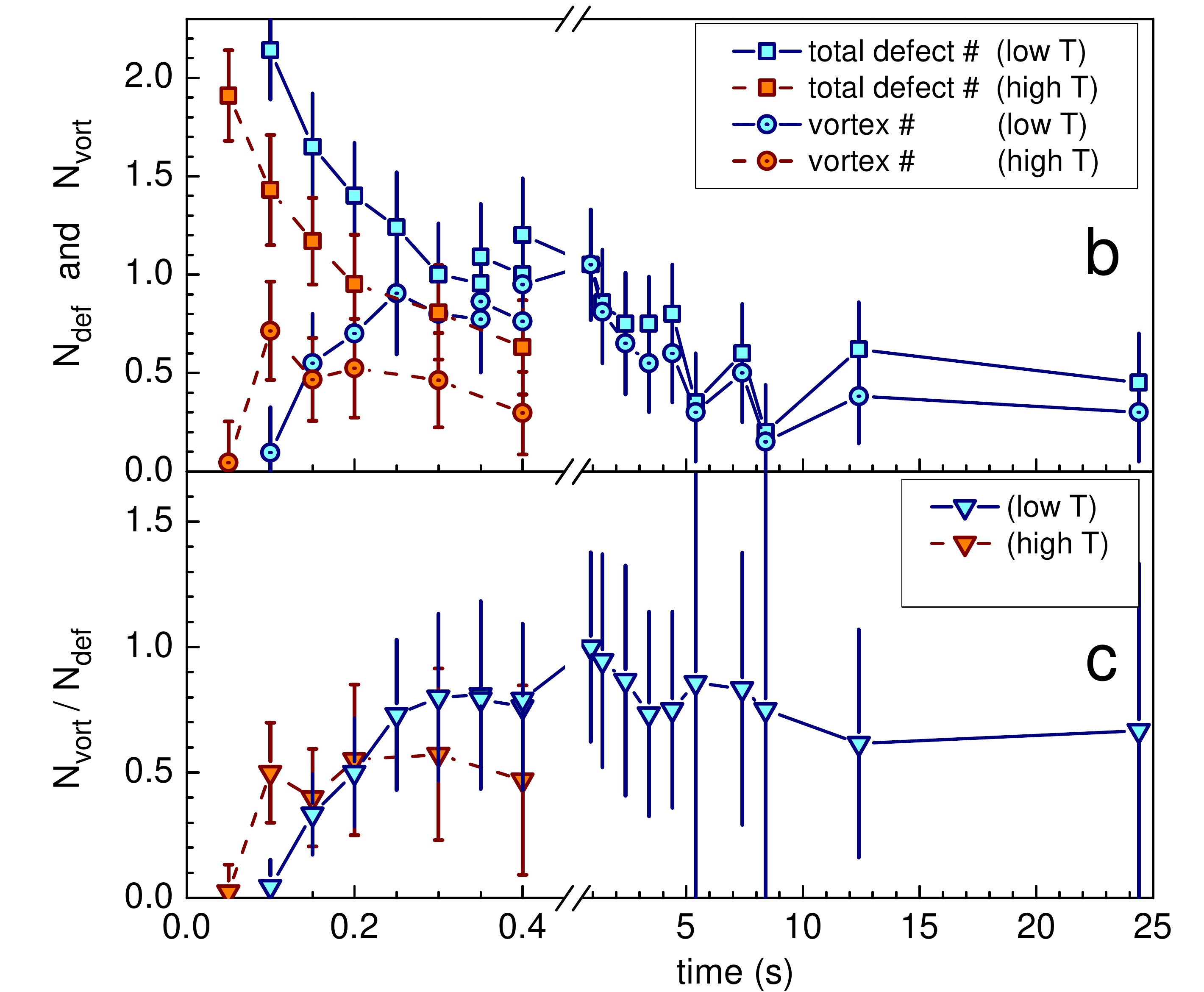}
\caption{(color online) a) Sketch of the evaporation ramp. b) Number of total defects (squares) and of solitonic vortices (dotted circles) measured on average at different times during and after the evaporation ramp. Blue (solid) and red (dashed) points correspond to datasets with different final evaporation frequencies. Error bars are evaluated as $\sqrt{\sigma^2+1/N_m}$, where $\sigma$ is the standard deviation of the average counts and $N_m$ the number of measurements. c) Ratio between averaged counted solitonic vortices and total defects.}
\label{fig:Figure4}
\end{figure}

In order to test this decay mechanism we probe our system close to the BEC transition, by interrupting the RF-evaporation ramp before its end, and we count the number of defects with a procedure similar to the one used in \cite{LamporesiKZM13}. In Fig.~\ref{fig:Figure4}b we give the average number of total defects (squares) found in each condensate as a function of the time after the occurrence of the BEC transition. On the same graph we also show the number of defects that we can safely identify as solitonic vortices (dotted circles) (i.e., those exhibiting a vortex line and/or a twist in the density dip). We also plot the ratio (Fig.~\ref{fig:Figure4}c) between average solitonic vortices and average total defects. Each point corresponds to the average over 20 images taken at a precise time during the evaporation ramp (left) or after a given waiting time following the end of the ramp (right). Two datasets are shown, corresponding to different final points of the evaporation ramp, as sketched in Fig.~\ref{fig:Figure4}a. Blue points (solid) are obtained by ramping down the temperature $T$ to low values, beyond our resolution, whereas red points (dashed) correspond to stopping evaporation shortly after the onset of condensation. The two datasets exhibit a similar trend, demonstrating that the initial fast variation of the defect number is scarcely dependent on the temperature (cooling) effect. We observe a decrease of the average number of defects with a decay time of about $\tau=90\;\text{ms}$. Within the same timescale, the fraction of defects identified as solitonic vortices grows from $0$ to almost $1$. Even though this growth can be partially attributed to an increase of the signal-to-noise ratio when moving away from the transition, the observed trend seems to be fully consistent with the presence of a decay channel of the defects created at the transition into solitonic vortices, some of them remaining in the condensates for many seconds \cite{note2}.
It is worth mentioning that several times we also observe pairs of vortex-antivortex lines \cite{SupplementalMaterial1} that are likely the two branches of a vortex ring developing into a solitonic vortex as predicted in \cite{Becker13,Reichl13,Wlazlowski14}.

To conclude, we reported on the experimental observations of solitonic vortices in a BEC of sodium atoms. We provide a full 3D imaging after free expansion, showing vortex filaments, vortex cores and twisted planar density depletions, as well as images of the phase dislocations associated to the quantized vorticity. We show how to distinguish vortices from antivortices by looking at the expanded density distribution and we finally provide quantitative data supporting the idea that solitonic vortices are likely the remnants of the snaking instability of grey solitons in elongated condensates. This work may serve as a basis for a systematic investigation of topological defects and solitons in quantum gases with variable geometry, from highly elongated quasi-1D to strongly oblate quasi-2D.

\begin{acknowledgments}
We thank Martin Zwierlein and Lawrence Cheuk for stimulating discussions and reciprocal exchange of unpublished data, and Andrea Invernizzi for technical assistance in the early stage of the experiment. We acknowledge financial support by ERC through the QGBE grant and the Provincia Autonoma di Trento. MT thanks the support of the PL-Grid Infrastructure.
\end{acknowledgments}

\newpage

\begin{figure*}[!hbt]
\centering
\includegraphics[width=1.9\columnwidth]{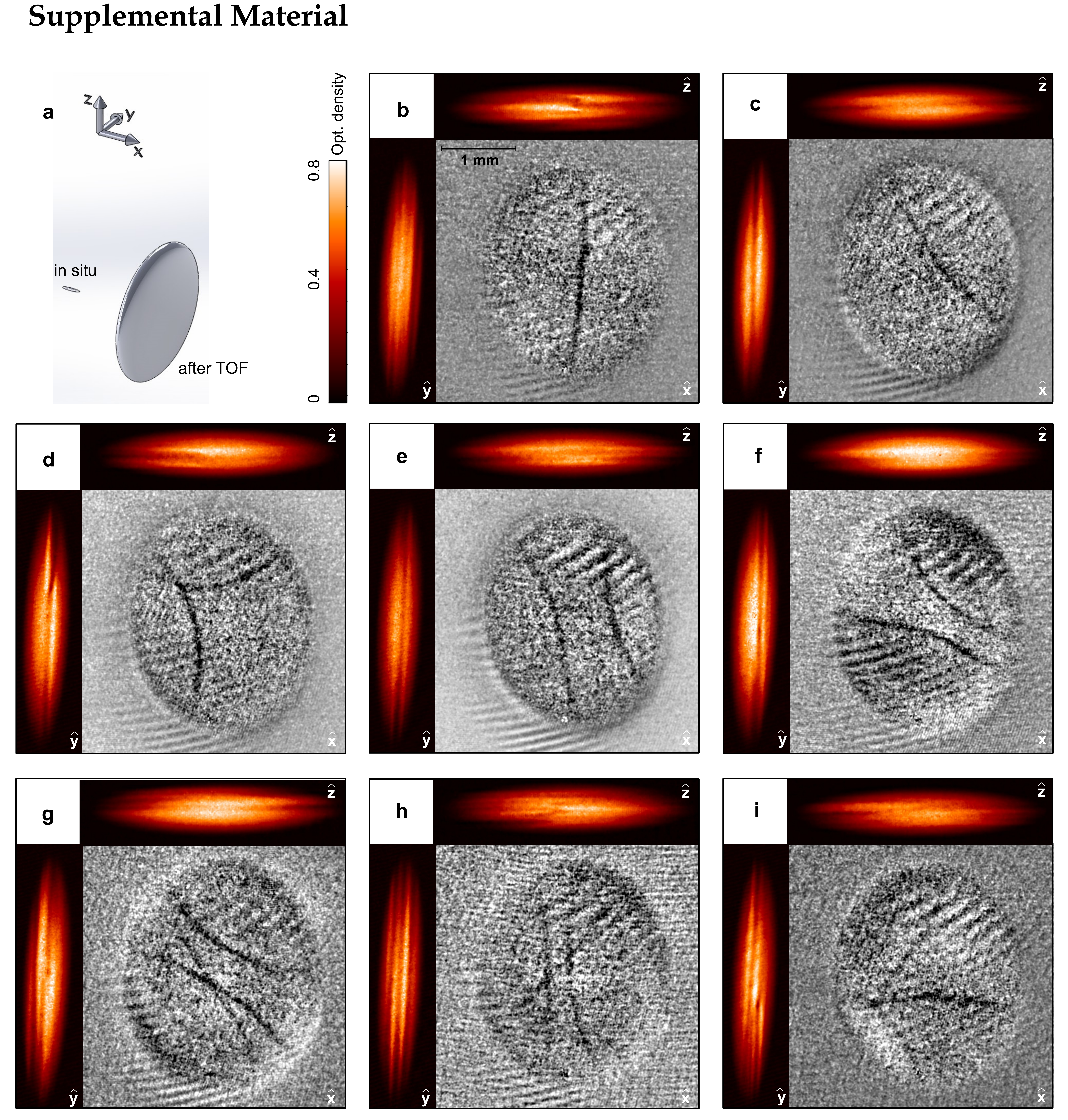}
\caption{(color online) Triaxial absorption images of an expanded condensate containing one or more solitonic vortices. One can notice that any time at least one end of the vortex line is aligned along one of the radial imaging axes ($y$ or $z$) the hollow vortex core is visible and the solitonic plane appears twisted around it. The vortex line can be straight or bent, but it always ends almost orthogonally to the condensate surface. Each vortex line lies in a radial plane due to its solitonic nature, therefore the radial imaging allows to reconstruct the axial position of most of the vortical lines.  a) Reference frame for the condensate with an elongated shape {\it in situ} and an oblate shape after long expansion. b) Straight solitonic vortex aligned along $z$ showing the twist when viewed from $z$ direction. c) Straight solitonic vortex not aligned along imaging axes and therefore appearing as a solitonic stripe without twists from $y$ and $z$. d-f) Pairs of vortex lines are often observed, probably reminiscent of the vortex ring that originated them \cite{Reichl13}. g-i) Examples with three vortex lines at different axial positions.}
\label{fig:FigureSM1}
\end{figure*}

\begin{figure*}[!hbt]
\centering
\includegraphics[width=1.85\columnwidth]{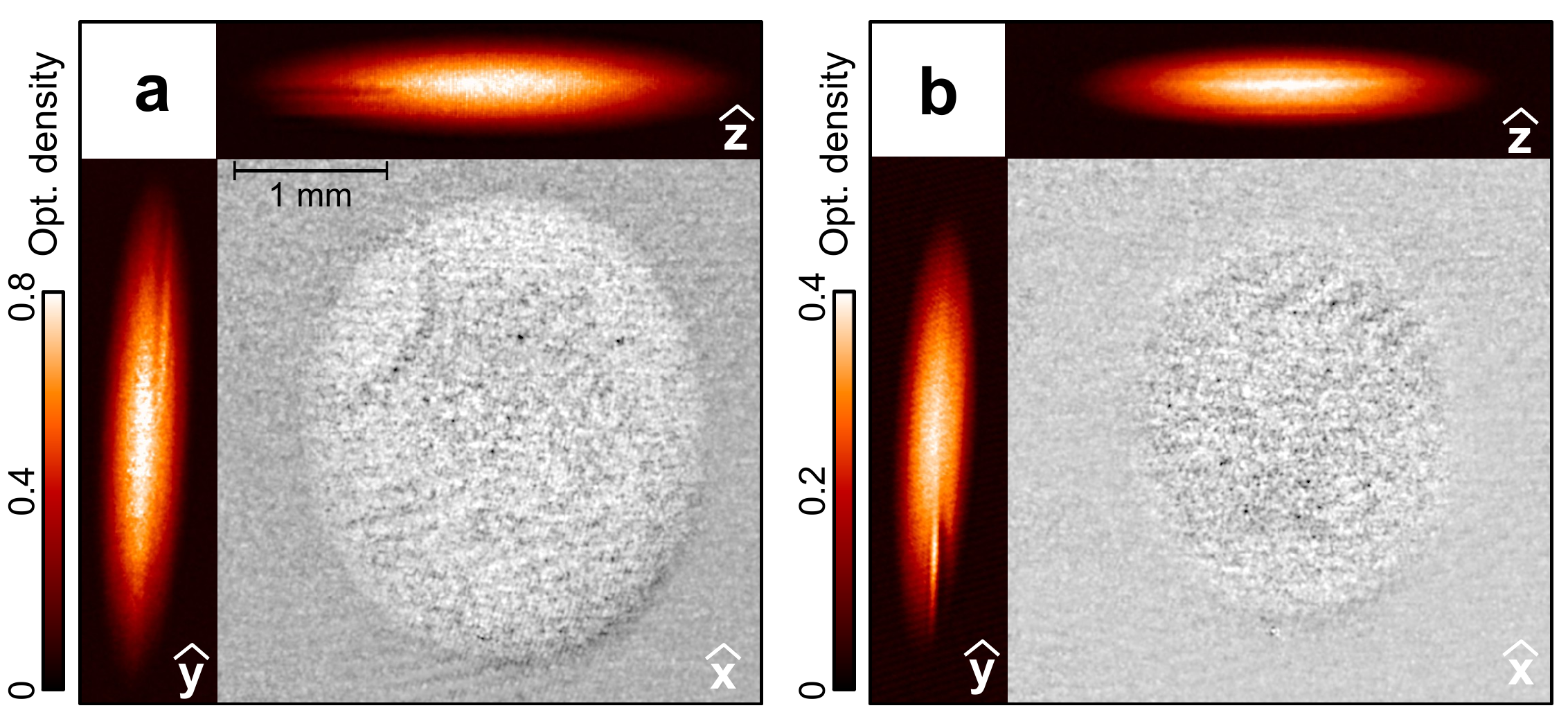}
\caption{(color online) Examples of triaxial imaging of thermalized condensates containing one defect at short and long waiting times. a) Set of images taken 400 ms after crossing the BEC transitions; this was counted as solitonic vortex because of the presence of the vortex arc on the residuals. b) Set of images taken after 12 seconds; this was counted as a general defect as it does not contain neither a twist in the density dip, nor a vortex line. Due to the finite lifetime of the condensate the atom number drops from 9 to 4 millions. }
\label{fig:FigureSM2}
\end{figure*}

\begin{figure*}[!hbt]
\centering
\includegraphics[width=1.9\columnwidth]{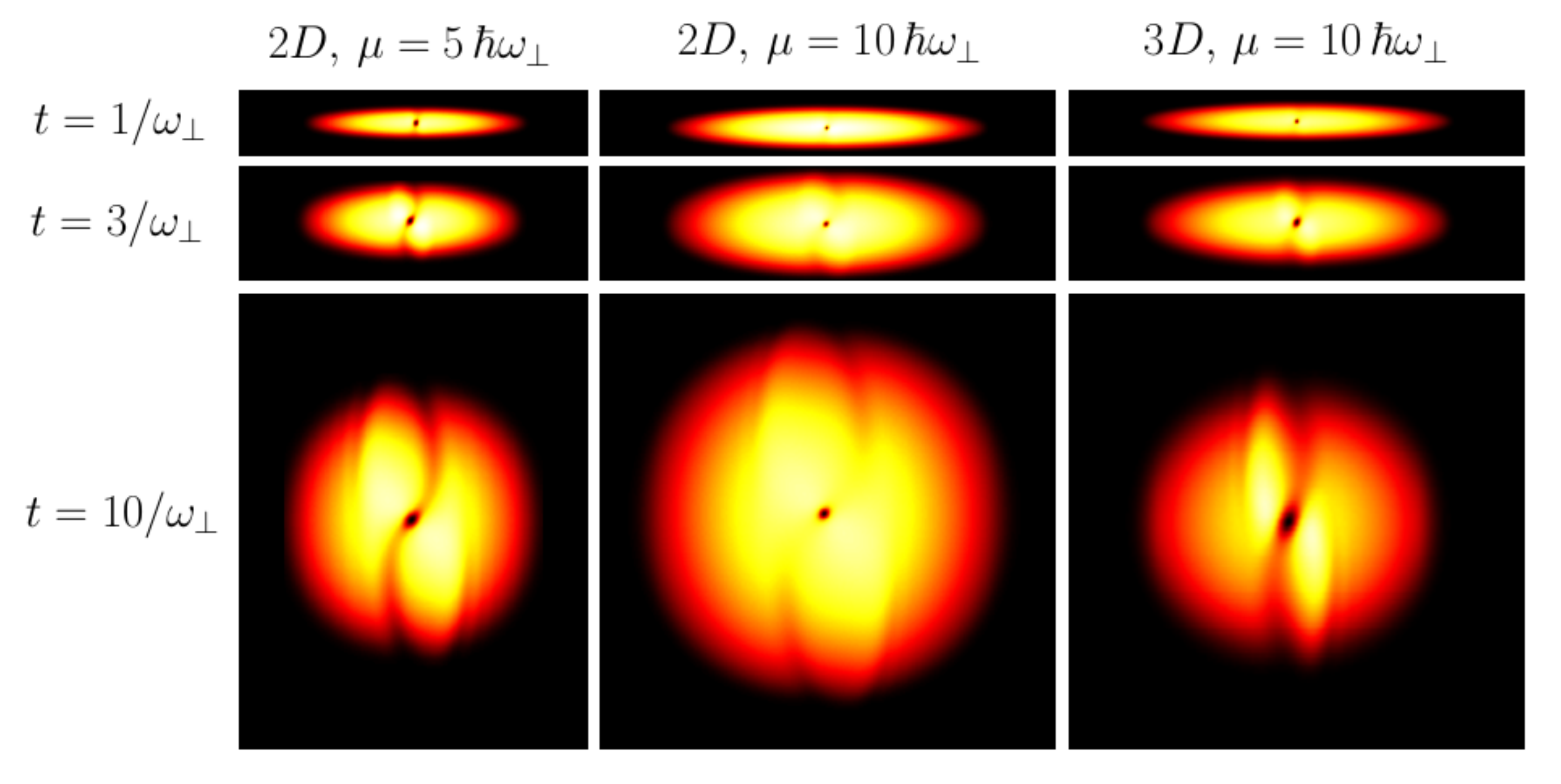}
\caption{(color online) The figure shows three instants of the BEC expansion obtained in GP simulations in two and three dimensions. In the first column, there is a 2D BEC cloud with chemical potential equal to $\mu = 5\, \hbar \omega_{\perp}$; in the second and third columns: 2D and 3D BECs respectively, both with $\mu = 10\, \hbar \omega_{\perp}$. For a given dimensionality, the larger the chemical potential $\mu$ the smaller the healing length and therefore the less visible solitonic depletions and vortex cores. In the 3D case, the density decreases faster, as the cloud expands also in the third dimension. Therefore, the healing length grows faster, which results in a wider depletion compared to the 2D with the same value of $\mu$. Also, in 3D, the system enters the ballistic regime earlier, so the density depletion twists less than in 2D.}
\label{fig:FigureSM3}
\end{figure*}


\begin{thebibliography}{10}%
\makeatletter
\providecommand \@ifxundefined [1]{%
 \ifx #1\undefined \expandafter \@firstoftwo
 \else \expandafter \@secondoftwo
\fi
}%
\providecommand \@ifnum [1]{%
 \ifnum #1\expandafter \@firstoftwo
 \else \expandafter \@secondoftwo
\fi
}%
\providecommand \enquote [1]{``#1''}%
\providecommand \bibnamefont  [1]{#1}%
\providecommand \bibfnamefont [1]{#1}%
\providecommand \citenamefont [1]{#1}%
\providecommand\href[0]{\@sanitize\@href}%
\providecommand\@href[1]{\endgroup\@@startlink{#1}\endgroup\@@href}%
\providecommand\@@href[1]{#1\@@endlink}%
\providecommand \@sanitize [0]{\begingroup\catcode`\&12\catcode`\#12\relax}%
\@ifxundefined \pdfoutput {\@firstoftwo}{%
 \@ifnum{\z@=\pdfoutput}{\@firstoftwo}{\@secondoftwo}%
}{%
 \providecommand\@@startlink[1]{\leavevmode}%
 \providecommand\@@endlink[0]{}%
}{%
 \providecommand\@@startlink[1]{%
  \leavevmode
  \pdfstartlink
   attr{/Border[0 0 1 ]/H/I/C[0 1 1]}%
   user{/Subtype/Link/A<</Type/Action/S/URI/URI(#1)>>}%
  \relax
 }%
 \providecommand\@@endlink[0]{\pdfendlink}%
}%
\providecommand \url  [0]{\begingroup\@sanitize \@url }%
\providecommand \@url [1]{\endgroup\@href {#1}{\urlprefix}}%
\providecommand \urlprefix [0]{URL }%
\providecommand \Eprint[0]{\href }%
\@ifxundefined \urlstyle {%
  \providecommand \doi [1]{doi:\discretionary{}{}{}#1}%
}{%
  \providecommand \doi [0]{doi:\discretionary{}{}{}\begingroup
  \urlstyle{rm}\Url }%
}%
\providecommand \doibase [0]{http://dx.doi.org/}%
\providecommand \Doi[1]{\href{\doibase#1}}%
\providecommand \bibAnnote [3]{%
  \BibitemShut{#1}%
  \begin{quotation}\noindent
    \textsc{Key:}\ #2\\\textsc{Annotation:}\ #3%
  \end{quotation}%
}%
\providecommand \bibAnnoteFile [2]{%
  \IfFileExists{#2}{\bibAnnote {#1} {#2} {\input{#2}}}{}%
}%
\providecommand \typeout [0]{\immediate \write \m@ne }%
\providecommand \selectlanguage [0]{\@gobble}%
\providecommand \bibinfo [0]{\@secondoftwo}%
\providecommand \bibfield [0]{\@secondoftwo}%
\providecommand \translation [1]{[#1]}%
\providecommand \BibitemOpen[0]{}%
\providecommand \bibitemStop [0]{}%
\providecommand \bibitemNoStop [0]{.\EOS\space}%
\providecommand \EOS [0]{\spacefactor3000\relax}%
\providecommand \BibitemShut [1]{\csname bibitem#1\endcsname}%
\bibitem{LamporesiKZM13}%
  \BibitemOpen
  \bibfield{author}{%
  \bibinfo {author} {\bibfnamefont{G.}~\bibnamefont{Lamporesi}}, \bibinfo
  {author} {\bibfnamefont{S.}~\bibnamefont{Donadello}}, \bibinfo {author}
  {\bibfnamefont{S.}~\bibnamefont{Serafini}}, \bibinfo {author}
  {\bibfnamefont{F.}~\bibnamefont{Dalfovo}},\ and\ \bibinfo {author}
  {\bibfnamefont{G.}~\bibnamefont{Ferrari}},\ }%
  \bibfield{journal}{%
  \bibinfo {journal} {Nature Phys.}\ }%
  \textbf{\bibinfo {volume} {9}},\ \bibinfo {pages} {656} (\bibinfo {year}
  {2013})%
  \bibAnnoteFile{NoStop}{LamporesiKZM13}%
\bibitem{Kibble76}%
  \BibitemOpen
  \bibfield{author}{%
  \bibinfo {author} {\bibfnamefont{T.~W.~B.}\ \bibnamefont{Kibble}},\ }%
  \bibfield{journal}{%
  \bibinfo {journal} {Journal of Physics A: Mathematical and General}\ }%
  \textbf{\bibinfo {volume} {9}},\ \bibinfo {pages} {1387} (\bibinfo {year}
  {1976})%
  \bibAnnoteFile{NoStop}{Kibble76}%
\bibitem{Zurek85}%
  \BibitemOpen
  \bibfield{author}{%
  \bibinfo {author} {\bibfnamefont{W.~H.}\ \bibnamefont{Zurek}},\ }%
  \bibfield{journal}{%
  \bibinfo {journal} {Nature}\ }%
  \textbf{\bibinfo {volume} {317}},\ \bibinfo {pages} {505} (\bibinfo {year}
  {1985})%
  \bibAnnoteFile{NoStop}{Zurek85}%
\bibitem{Zurek09}%
  \BibitemOpen
  \bibfield{author}{%
  \bibinfo {author} {\bibfnamefont{W.~H.}\ \bibnamefont{Zurek}},\ }%
  \bibfield{journal}{%
  \Doi{10.1103/PhysRevLett.102.105702}{\bibinfo {journal} {Phys. Rev. Lett.}}\
  }%
  \textbf{\bibinfo {volume} {102}},\ \bibinfo {pages} {105702} (\bibinfo {year}
  {2009})%
  \bibAnnoteFile{NoStop}{Zurek09}%
\bibitem{Feder00}%
  \BibitemOpen
  \bibfield{author}{%
  \bibinfo {author} {\bibfnamefont{D.~L.}\ \bibnamefont{Feder}}, \bibinfo
  {author} {\bibfnamefont{M.~S.}\ \bibnamefont{Pindzola}}, \bibinfo {author}
  {\bibfnamefont{L.~A.}\ \bibnamefont{Collins}}, \bibinfo {author}
  {\bibfnamefont{B.~I.}\ \bibnamefont{Schneider}},\ and\ \bibinfo {author}
  {\bibfnamefont{C.~W.}\ \bibnamefont{Clark}},\ }%
  \bibfield{journal}{%
  \Doi{10.1103/PhysRevA.62.053606}{\bibinfo {journal} {Phys. Rev. A}}\ }%
  \textbf{\bibinfo {volume} {62}},\ \bibinfo {pages} {053606} (\bibinfo {year}
  {2000})%
  \bibAnnoteFile{NoStop}{Feder00}%
\bibitem{Yefsah13}%
  \BibitemOpen
  \bibfield{author}{%
  \bibinfo {author} {\bibfnamefont{T.}~\bibnamefont{Yefsah}}, \bibinfo {author}
  {\bibfnamefont{A.~T.}\ \bibnamefont{Sommer}}, \bibinfo {author}
  {\bibfnamefont{M.~J.~H.}\ \bibnamefont{Ku}}, \bibinfo {author}
  {\bibfnamefont{L.~W.}\ \bibnamefont{Cheuk}}, \bibinfo {author}
  {\bibfnamefont{W.}~\bibnamefont{Ji}}, \bibinfo {author}
  {\bibfnamefont{W.~S.}\ \bibnamefont{Bakr}},\ and\ \bibinfo {author}
  {\bibfnamefont{M.~W.}\ \bibnamefont{Zwierlein}},\ }%
  \bibfield{journal}{%
  \bibinfo {journal} {Nature}\ }%
  \textbf{\bibinfo {volume} {499}},\ \bibinfo {pages} {426} (\bibinfo {year}
  {2013})%
  \bibAnnoteFile{NoStop}{Yefsah13}%
\bibitem{Brand02}%
  \BibitemOpen
  \bibfield{author}{%
  \bibinfo {author} {\bibfnamefont{J.}~\bibnamefont{Brand}}\ and\ \bibinfo
  {author} {\bibfnamefont{W.~P.}\ \bibnamefont{Reinhardt}},\ }%
  \bibfield{journal}{%
  \Doi{10.1103/PhysRevA.65.043612}{\bibinfo {journal} {Phys. Rev. A}}\ }%
  \textbf{\bibinfo {volume} {65}},\ \bibinfo {pages} {043612} (\bibinfo {year}
  {2002})%
  \bibAnnoteFile{NoStop}{Brand02}%
\bibitem{Komineas03}%
  \BibitemOpen
  \bibfield{author}{%
  \bibinfo {author} {\bibfnamefont{S.}~\bibnamefont{Komineas}}\ and\ \bibinfo
  {author} {\bibfnamefont{N.}~\bibnamefont{Papanicolaou}},\ }%
  \bibfield{journal}{%
  \Doi{10.1103/PhysRevA.68.043617}{\bibinfo {journal} {Phys. Rev. A}}\ }%
  \textbf{\bibinfo {volume} {68}},\ \bibinfo {pages} {043617} (\bibinfo {year}
  {2003})%
  \bibAnnoteFile{NoStop}{Komineas03}%
\bibitem{Becker13}%
  \BibitemOpen
  \bibfield{author}{%
  \bibinfo {author} {\bibfnamefont{C.}~\bibnamefont{Becker}}, \bibinfo {author}
  {\bibfnamefont{K.}~\bibnamefont{Sengstock}}, \bibinfo {author}
  {\bibfnamefont{P.}~\bibnamefont{Schmelcher}}, \bibinfo {author}
  {\bibfnamefont{P.~G.}\ \bibnamefont{Kevrekidis}},\ and\ \bibinfo {author}
  {\bibfnamefont{R.}~\bibnamefont{Carretero-Gonz{\'a}lez}},\ }%
  \bibfield{journal}{%
  \bibinfo {journal} {New J. Phys.}\ }%
  \textbf{\bibinfo {volume} {15}},\ \bibinfo {pages} {113028} (\bibinfo {year}
  {2013})%
  \bibAnnoteFile{NoStop}{Becker13}%
\bibitem{Ku14}%
  \BibitemOpen
  \bibfield{author}{%
  \bibinfo {author} {\bibfnamefont{M.}~\bibnamefont{Ku}}, \bibinfo {author}
  {\bibfnamefont{W.}~\bibnamefont{Ji}}, \bibinfo {author}
  {\bibfnamefont{B.}~\bibnamefont{Mukherjee}}, \bibinfo {author}
  {\bibfnamefont{E.}~\bibnamefont{Guarado-Sanchez}}, \bibinfo {author}
  {\bibfnamefont{L.}~\bibnamefont{Cheuk}}, \bibinfo {author}
  {\bibfnamefont{T.}~\bibnamefont{Yefsah}},\ and\ \bibinfo {author}
  {\bibfnamefont{M.}~\bibnamefont{Zwierlein}},\ }%
  \bibinfo {journal} {arXiv:1402.7052}%
  \bibAnnoteFile{NoStop}{Ku14}%
\bibitem{Reichl13}%
  \BibitemOpen
\bibfield{journal}{%
    }%
  \bibfield{author}{%
  \bibinfo {author} {\bibfnamefont{M.~D.}\ \bibnamefont{Reichl}}\ and\ \bibinfo
  {author} {\bibfnamefont{E.~J.}\ \bibnamefont{Mueller}},\ }%
  \bibfield{journal}{%
  \Doi{10.1103/PhysRevA.88.053626}{\bibinfo {journal} {Phys. Rev. A}}\ }%
  \textbf{\bibinfo {volume} {88}},\ \bibinfo {pages} {053626} (\bibinfo {year}
  {2013})%
  \bibAnnoteFile{NoStop}{Reichl13}%
\bibitem{Bulgac14}%
  \BibitemOpen
  \bibfield{author}{%
  \bibinfo {author} {\bibfnamefont{A.}~\bibnamefont{Bulgac}}, \bibinfo {author}
  {\bibfnamefont{M.}\ \bibnamefont{McNeil~Forbes}}, \bibinfo {author}
  {\bibfnamefont{M.~M.}\ \bibnamefont{Kelley}}, \bibinfo {author}
  {\bibfnamefont{K.~J.}\ \bibnamefont{Roche}},\ and\ \bibinfo {author}
  {\bibfnamefont{G.}~\bibnamefont{Wlaz\l{}owski}},\ }%
  \bibfield{journal}{%
  \Doi{10.1103/PhysRevLett.112.025301}{\bibinfo {journal} {Phys. Rev. Lett.}}\
  }%
  \textbf{\bibinfo {volume} {112}},\ \bibinfo {pages} {025301} (\bibinfo {year}
  {2014})%
  \bibAnnoteFile{NoStop}{Bulgac14}%
\bibitem{Wlazlowski14}%
  \BibitemOpen
  \bibfield{author}{%
  \bibinfo {author} {\bibfnamefont{G.}~\bibnamefont{Wlaz\l{}owski}}, \bibinfo
  {author} {\bibfnamefont{A.}~\bibnamefont{Bulgac}}, \bibinfo {author}
  {\bibfnamefont{M.}~\bibnamefont{McNeil~Forbes}},\ and\ \bibinfo {author}
  {\bibfnamefont{K.}~\bibnamefont{Roche}},\ }%
  \bibinfo {journal} {arXiv:1404.1038v1}%
  \bibAnnoteFile{NoStop}{Wlazlowski14}%
\bibitem{Scherpelz14}%
  \BibitemOpen
\bibfield{journal}{%
    }%
  \bibfield{author}{%
  \bibinfo {author} {\bibfnamefont{P.}~\bibnamefont{Scherpelz}}, \bibinfo
  {author} {\bibfnamefont{K.}~\bibnamefont{Padavi{\'c}}}, \bibinfo {author}
  {\bibfnamefont{A.}~\bibnamefont{Ran\c{c}on}}, \bibinfo {author}
  {\bibfnamefont{A.}~\bibnamefont{Glatz}}, \bibinfo {author}
  {\bibfnamefont{I.}~\bibnamefont{Aranson}},\ and\ \bibinfo {author}
  {\bibfnamefont{K.}~\bibnamefont{Levin}},\ }%
  \bibinfo {journal} {arXiv:1401.8267}%
  \bibAnnoteFile{NoStop}{Scherpelz14}%
\bibitem{Lamporesi13}%
  \BibitemOpen
\bibfield{journal}{%
    }%
  \bibfield{author}{%
  \bibinfo {author} {\bibfnamefont{G.}~\bibnamefont{Lamporesi}}, \bibinfo
  {author} {\bibfnamefont{S.}~\bibnamefont{Donadello}}, \bibinfo {author}
  {\bibfnamefont{S.}~\bibnamefont{Serafini}},\ and\ \bibinfo {author}
  {\bibfnamefont{G.}~\bibnamefont{Ferrari}},\ }%
  \bibfield{journal}{%
  \bibinfo {journal} {Rev. Sci. Instrum.}\ }%
  \textbf{\bibinfo {volume} {84}},\ \bibinfo {pages} {063102} (\bibinfo {year}
  {2013})%
  \bibAnnoteFile{NoStop}{Lamporesi13}%
\bibitem{SupplementalMaterial1}%
  \BibitemOpen
  \bibinfo {title} {\rm See Supplemental Material at URL for images
  of detected solitonic vortices and for 2D and 3D simulations.}\ %
\bibitem{Chevy00}%
  \BibitemOpen
  \bibfield{author}{%
  \bibinfo {author} {\bibfnamefont{F.}~\bibnamefont{Chevy}}, \bibinfo {author}
  {\bibfnamefont{K.~W.}\ \bibnamefont{Madison}},\ and\ \bibinfo {author}
  {\bibfnamefont{J.}~\bibnamefont{Dalibard}},\ }%
  \bibfield{journal}{%
  \Doi{10.1103/PhysRevLett.85.2223}{\bibinfo {journal} {Phys. Rev. Lett.}}\ }%
  \textbf{\bibinfo {volume} {85}},\ \bibinfo {pages} {2223} (\bibinfo {year}
  {2000})%
  \bibAnnoteFile{NoStop}{Chevy00}%
\bibitem{Anderson2000}%
  \BibitemOpen
  \bibfield{author}{%
  \bibinfo {author} {\bibfnamefont{B.~P.}\ \bibnamefont{Anderson}}, \bibinfo
  {author} {\bibfnamefont{P.~C.}\ \bibnamefont{Haljan}}, \bibinfo {author}
  {\bibfnamefont{C.~E.}\ \bibnamefont{Wieman}},\ and\ \bibinfo {author}
  {\bibfnamefont{E.~A.}\ \bibnamefont{Cornell}},\ }%
  \bibfield{journal}{%
  \Doi{10.1103/PhysRevLett.85.2857}{\bibinfo {journal} {Phys. Rev. Lett.}}\ }%
  \textbf{\bibinfo {volume} {85}},\ \bibinfo {pages} {2857} (\bibinfo {year}
  {2000})%
  \bibAnnoteFile{NoStop}{Anderson2000}%
\bibitem{Freilich10}%
  \BibitemOpen
  \bibfield{author}{%
  \bibinfo {author} {\bibfnamefont{D.~V.}\ \bibnamefont{Freilich}}, \bibinfo
  {author} {\bibfnamefont{D.~M.}\ \bibnamefont{Bianchi}}, \bibinfo {author}
  {\bibfnamefont{A.~M.}\ \bibnamefont{Kaufman}}, \bibinfo {author}
  {\bibfnamefont{T.~K.}\ \bibnamefont{Langin}},\ and\ \bibinfo {author}
  {\bibfnamefont{D.~S.}\ \bibnamefont{Hall}},\ }%
  \bibfield{journal}{%
  \bibinfo {journal} {Science}\ }%
  \textbf{\bibinfo {volume} {329}},\ \bibinfo {pages} {1182} (\bibinfo {year}
  {2010})%
  \bibAnnoteFile{NoStop}{Freilich10}%
\bibitem{Chevy01}%
  \BibitemOpen
  \bibfield{author}{%
  \bibinfo {author} {\bibfnamefont{F.}~\bibnamefont{Chevy}}, \bibinfo {author}
  {\bibfnamefont{K.~W.}\ \bibnamefont{Madison}}, \bibinfo {author}
  {\bibfnamefont{V.}~\bibnamefont{Bretin}},\ and\ \bibinfo {author}
  {\bibfnamefont{J.}~\bibnamefont{Dalibard}},\ }%
  \bibfield{journal}{%
  \Doi{10.1103/PhysRevA.64.031601}{\bibinfo {journal} {Phys. Rev. A}}\ }%
  \textbf{\bibinfo {volume} {64}},\ \bibinfo {pages} {031601} (\bibinfo {year}
  {2001})%
  \bibAnnoteFile{NoStop}{Chevy01}%
\bibitem{ParkerPhDthesis}%
  \BibitemOpen
  \bibfield{author}{%
  \bibinfo {author} {\bibfnamefont{N.}~\bibnamefont{Parker}},\ }%
  \emph{\bibinfo {title} {Numerical Studies of Vortices and Dark Solitons in
  Atomic Bose-Einstein Condensates}},\ Ph.D. thesis,\ \bibinfo {school}
  {University of Durham} (\bibinfo {year} {2004})%
  \bibAnnoteFile{NoStop}{ParkerPhDthesis}%
\bibitem{Castin96}%
  \BibitemOpen
  \bibfield{author}{%
  \bibinfo {author} {\bibfnamefont{Y.}~\bibnamefont{Castin}}\ and\ \bibinfo
  {author} {\bibfnamefont{R.}~\bibnamefont{Dum}},\ }%
  \bibfield{journal}{%
  \Doi{10.1103/PhysRevLett.77.5315}{\bibinfo {journal} {Phys. Rev. Lett.}}\ }%
  \textbf{\bibinfo {volume} {77}},\ \bibinfo {pages} {5315} (\bibinfo {year}
  {1996})%
  \bibAnnoteFile{NoStop}{Castin96}%
\bibitem{Kagan96}%
  \BibitemOpen
  \bibfield{author}{%
  \bibinfo {author} {\bibfnamefont{Y.}~\bibnamefont{Kagan}}, \bibinfo {author}
  {\bibfnamefont{E.~L.}\ \bibnamefont{Surkov}},\ and\ \bibinfo {author}
  {\bibfnamefont{G.~V.}\ \bibnamefont{Shlyapnikov}},\ }%
  \bibfield{journal}{%
  \Doi{10.1103/PhysRevA.54.R1753}{\bibinfo {journal} {Phys. Rev. A}}\ }%
  \textbf{\bibinfo {volume} {54}},\ \bibinfo {pages} {R1753} (\bibinfo {year}
  {1996})%
  \bibAnnoteFile{NoStop}{Kagan96}%
\bibitem{Dalfovo97}%
  \BibitemOpen
  \bibfield{author}{%
  \bibinfo {author} {\bibfnamefont{F.}~\bibnamefont{Dalfovo}}, \bibinfo
  {author} {\bibfnamefont{C.}~\bibnamefont{Minniti}}, \bibinfo {author}
  {\bibfnamefont{S.}~\bibnamefont{Stringari}},\ and\ \bibinfo {author}
  {\bibfnamefont{L.}~\bibnamefont{Pitaevskii}},\ }%
  \bibfield{journal}{%
  \Doi{, publisher = {}}{\bibinfo {journal} {Phys. Lett. A}}\ }%
  \textbf{\bibinfo {volume} {227}},\ \bibinfo {pages} {259} (\bibinfo {year}
  {1997})%
  \bibAnnoteFile{NoStop}{Dalfovo97}%
\bibitem{Modugno03}%
  \BibitemOpen
  \bibfield{author}{%
  \bibinfo {author} {\bibfnamefont{M.}~\bibnamefont{Modugno}}, \bibinfo
  {author} {\bibfnamefont{L.}~\bibnamefont{Pricoupenko}},\ and\ \bibinfo
  {author} {\bibfnamefont{Y.}~\bibnamefont{Castin}},\ }%
  \bibfield{journal}{%
  \Doi{10.1140/epjd/e2003-00015-y}{\bibinfo {journal} {Eur. Phys. J. D}}\ }%
  \textbf{\bibinfo {volume} {22}},\ \bibinfo {pages} {235} (\bibinfo {year}
  {2003})%
  \bibAnnoteFile{NoStop}{Modugno03}%
\bibitem{Massignan03}%
  \BibitemOpen
  \bibfield{author}{%
  \bibinfo {author} {\bibfnamefont{P.}~\bibnamefont{Massignan}}\ and\ \bibinfo
  {author} {\bibfnamefont{M.}~\bibnamefont{Modugno}},\ }%
  \bibfield{journal}{%
  \Doi{10.1103/PhysRevA.67.023614}{\bibinfo {journal} {Phys. Rev. A}}\ }%
  \textbf{\bibinfo {volume} {67}},\ \bibinfo {pages} {023614} (\bibinfo {year}
  {2003})%
  \bibAnnoteFile{NoStop}{Massignan03}%
\bibitem{Tozzo04}%
  \BibitemOpen
  \bibfield{author}{%
  \bibinfo {author} {\bibfnamefont{C.}~\bibnamefont{Tozzo}}\ and\ \bibinfo
  {author} {\bibfnamefont{F.}~\bibnamefont{Dalfovo}},\ }%
  \bibfield{journal}{%
  \bibinfo {journal} {Phys. Rev. A}\ }%
  \textbf{\bibinfo {volume} {69}},\ \bibinfo {pages} {053606} (\bibinfo {year}
  {2004})%
  \bibAnnoteFile{NoStop}{Tozzo04}%
\bibitem{Dalfovo99}%
  \BibitemOpen
  \bibfield{author}{%
  \bibinfo {author} {\bibfnamefont{F.}~\bibnamefont{Dalfovo}}, \bibinfo
  {author} {\bibfnamefont{S.}~\bibnamefont{Giorgini}}, \bibinfo {author}
  {\bibfnamefont{L.~P.}\ \bibnamefont{Pitaevskii}},\ and\ \bibinfo {author}
  {\bibfnamefont{S.}~\bibnamefont{Stringari}},\ }%
  \bibfield{journal}{%
  \Doi{10.1103/RevModPhys.71.463}{\bibinfo {journal} {Rev. Mod. Phys.}}\ }%
  \textbf{\bibinfo {volume} {71}},\ \bibinfo {pages} {463} (\bibinfo {year}
  {1999})%
  \bibAnnoteFile{NoStop}{Dalfovo99}%
\bibitem{note1}%
  \BibitemOpen
  \bibinfo {title} {\rm Simulations with larger values of the
  chemical potential are beyond the limits of our computational resources.}\
  %
\bibitem{Bolda98}%
  \BibitemOpen
  \bibfield{author}{%
  \bibinfo {author} {\bibfnamefont{E.~L.}\ \bibnamefont{Bolda}}\ and\ \bibinfo
  {author} {\bibfnamefont{D.~F.}\ \bibnamefont{Walls}},\ }%
  \bibfield{journal}{%
  \Doi{10.1103/PhysRevLett.81.5477}{\bibinfo {journal} {Phys. Rev. Lett.}}\ }%
  \textbf{\bibinfo {volume} {81}},\ \bibinfo {pages} {5477} (\bibinfo {year}
  {1998})%
  \bibAnnoteFile{NoStop}{Bolda98}%
\bibitem{Inouye01}%
  \BibitemOpen
  \bibfield{author}{%
  \bibinfo {author} {\bibfnamefont{S.}~\bibnamefont{Inouye}}, \bibinfo {author}
  {\bibfnamefont{S.}~\bibnamefont{Gupta}}, \bibinfo {author}
  {\bibfnamefont{T.}~\bibnamefont{Rosenband}}, \bibinfo {author}
  {\bibfnamefont{A.~P.}\ \bibnamefont{Chikkatur}}, \bibinfo {author}
  {\bibfnamefont{A.}~\bibnamefont{G\"orlitz}}, \bibinfo {author}
  {\bibfnamefont{T.~L.}\ \bibnamefont{Gustavson}}, \bibinfo {author}
  {\bibfnamefont{A.~E.}\ \bibnamefont{Leanhardt}}, \bibinfo {author}
  {\bibfnamefont{D.~E.}\ \bibnamefont{Pritchard}},\ and\ \bibinfo {author}
  {\bibfnamefont{W.}~\bibnamefont{Ketterle}},\ }%
  \bibfield{journal}{%
  \Doi{10.1103/PhysRevLett.87.080402}{\bibinfo {journal} {Phys. Rev. Lett.}}\
  }%
  \textbf{\bibinfo {volume} {87}},\ \bibinfo {pages} {080402} (\bibinfo {year}
  {2001})%
  \bibAnnoteFile{NoStop}{Inouye01}%
\bibitem{Kozuma99}%
  \BibitemOpen
  \bibfield{author}{%
  \bibinfo {author} {\bibfnamefont{M.}~\bibnamefont{Kozuma}}, \bibinfo {author}
  {\bibfnamefont{L.}~\bibnamefont{Deng}}, \bibinfo {author}
  {\bibfnamefont{E.~W.}\ \bibnamefont{Hagley}}, \bibinfo {author}
  {\bibfnamefont{J.}~\bibnamefont{Wen}}, \bibinfo {author}
  {\bibfnamefont{R.}~\bibnamefont{Lutwak}}, \bibinfo {author}
  {\bibfnamefont{K.}~\bibnamefont{Helmerson}}, \bibinfo {author}
  {\bibfnamefont{S.~L.}\ \bibnamefont{Rolston}},\ and\ \bibinfo {author}
  {\bibfnamefont{W.~D.}\ \bibnamefont{Phillips}},\ }%
  \bibfield{journal}{%
  \Doi{10.1103/PhysRevLett.82.871}{\bibinfo {journal} {Phys. Rev. Lett.}}\ }%
  \textbf{\bibinfo {volume} {82}},\ \bibinfo {pages} {871} (\bibinfo {year}
  {1999})%
  \bibAnnoteFile{NoStop}{Kozuma99}%
\bibitem{Shomroni09}%
  \BibitemOpen
  \bibfield{author}{%
  \bibinfo {author} {\bibfnamefont{I.}~\bibnamefont{Shomroni}}, \bibinfo
  {author} {\bibfnamefont{E.}~\bibnamefont{Lahoud}}, \bibinfo {author}
  {\bibfnamefont{S.}~\bibnamefont{Levy}},\ and\ \bibinfo {author}
  {\bibfnamefont{J.}~\bibnamefont{Steinhauer}},\ }%
  \bibfield{journal}{%
  \Doi{10.1038/nphys1177}{\bibinfo {journal} {Nature Physics}}\ }%
  \textbf{\bibinfo {volume} {5}},\ \bibinfo {pages} {193} (\bibinfo {year}
  {2009})%
  \bibAnnoteFile{NoStop}{Shomroni09}%
\bibitem{Kamchatnov08}%
  \BibitemOpen
  \bibfield{author}{%
  \bibinfo {author} {\bibfnamefont{A.~M.}\ \bibnamefont{Kamchatnov}}\ and\
  \bibinfo {author} {\bibfnamefont{L.~P.}\ \bibnamefont{Pitaevskii}},\ }%
  \bibfield{journal}{%
  \Doi{10.1103/PhysRevLett.100.160402}{\bibinfo {journal} {Phys. Rev. Lett.}}\
  }%
  \textbf{\bibinfo {volume} {100}},\ \bibinfo {pages} {160402} (\bibinfo {year}
  {2008})%
  \bibAnnoteFile{NoStop}{Kamchatnov08}%
\bibitem{Cetoli13}%
  \BibitemOpen
  \bibfield{author}{%
  \bibinfo {author} {\bibfnamefont{A.}~\bibnamefont{Cetoli}}, \bibinfo {author}
  {\bibfnamefont{J.}~\bibnamefont{Brand}}, \bibinfo {author}
  {\bibfnamefont{R.~G.}\ \bibnamefont{Scott}}, \bibinfo {author}
  {\bibfnamefont{F.}~\bibnamefont{Dalfovo}},\ and\ \bibinfo {author}
  {\bibfnamefont{L.~P.}\ \bibnamefont{Pitaevskii}},\ }%
  \bibfield{journal}{%
  \Doi{10.1103/PhysRevA.88.043639}{\bibinfo {journal} {Phys. Rev. A}}\ }%
  \textbf{\bibinfo {volume} {88}},\ \bibinfo {pages} {043639} (\bibinfo {year}
  {2013})%
  \bibAnnoteFile{NoStop}{Cetoli13}%
\bibitem{note2}%
  \BibitemOpen
  \bibinfo {title} {\rm Although not all defects are identified as
  solitonic vortices we expect them all to contain vorticity after several
  seconds of wait time. A large tilt of the vortex line with respect to the
  imaging axes and edge position of the defect along the axial coordinate of
  the condensate \protect\cite{SupplementalMaterial1} might contribute in
  masking their nature.}\ %
\end{thebibliography}
\end{document}